\documentclass[rnote]{aa} 
\usepackage[authoryear]{natbib}
\bibpunct{(}{)}{;}{a}{}{,} 

\authorrunning{Orozco Su\'arez et al.\/} 
\titlerunning{Simultaneous ME inversion of the \ion{Fe}{I} 630~nm lines} 

\usepackage{graphicx}
\usepackage{txfonts}
\usepackage{color}

\begin{document}
   \title{Milne-Eddington inversion of the Fe I line pair at 630~nm}

   \author{D.\ Orozco Su\'arez \inst{1}, L.~R.\ Bellot Rubio \inst{2}, 
   \and J.~C. del Toro Iniesta\inst{2}}

  \institute{National Astronomical Observatory of Japan, 2-21-1 Osawa, 
             Mitaka, Tokyo 181-8588, Japan, \email{d.orozco@nao.ac.jp}
   \and 
 	Instituto de Astrof\'{\i}sica de Andaluc\'{\i}a (CSIC), 
        Apdo.\ Correos 3004, 18080 Granada, Spain}

   \date{Received ; accepted }

\abstract
{The iron lines at 630.15 and 630.25~nm are often used to determine
the physical conditions of the solar photosphere. A common approach is
to invert them simultaneously under the Milne-Eddington
approximation. The same thermodynamic parameters are employed for the
two lines, except for their opacities, which are assumed to have a
constant ratio. } {We aim at investigating the validity of this
assumption, since the two lines are not exactly the same.} {We use
magnetohydrodynamic simulations of the quiet Sun to examine the
behavior of the ME thermodynamic parameters and their influence on the
retrieval of vector magnetic fields and flow velocities.}  {Our
analysis shows that the two lines can be coupled and inverted
simultaneously using the same thermodynamic parameters and a constant
opacity ratio. The inversion of two lines is significantly more
accurate than single-line inversions because of the larger number of
observables. } {}


 \maketitle

%

 \section{Introduction} 
  \label{sec:intro}
 
Both solar physics and radiative transfer owe many things to the
Milne-Eddington (ME) approximation. Being an incredibly simplistic
description of spectral line formation, it provides useful hints to
understand the behavior of spectral lines as formed in solar and
stellar atmospheres. It also offers a key diagnostics to infer the
physical conditions of the plasma. This is particularly true when
magnetic fields are present. The first solution of the radiative
transfer equation in the presence of a magnetic field was derived
adopting the ME assumption that all physical quantities relevant to
line formation are constant with depth
\citep{1956PASJ....8..108U,1962aIzvKrAO...27.148R,1962aIzvKrAO...28.259R}. 
Under these conditions, the solution is analytic and, therefore, by
simply varying the model parameters, one gets a handle on the behavior
of Stokes line profiles. Similarly, the analytic character of the ME
solution allows perturbative analyses like those performed by
\cite{1983SoPh...87..221L}, who studied the influence of velocity
gradients, or \cite{orozco2007}, who calculated the
sensitivities of ME Stokes profiles to the various model
parameters. Inversion codes of the radiative transfer equation are
diagnostic tools that become simpler under the ME hypothesis. Since
the pioneering work by \cite{1972lfpm.conf..227H} and 
\cite{1977SoPh...55...47A}, and after improvements by 
\cite{1982SoPh...78..355L} and \cite{1984SoPh...93..269L}, a 
number of ME inversion codes have been developed. These include the
HAO code \citep{lites2,1990ApJ...348..747L}, MELANIE
\citep{socas2001}, HELIX \citep{2004A&A...414.1109L}, MILOS
\citep{orozco2007}, and VFISV \citep{2010SoPh..tmp...35B}.

Strictly speaking, the ME model is applicable to just one spectral
line. The reason is that the so-called thermodynamic parameters of the
model are meant to characterize the behavior of the specific line
under consideration. The line-to-continuum opacity ratio, $\eta_{0}$,
the Doppler width of the line, $\Delta \lambda_{\rm D}$, and the
damping parameter, $a$, govern the shape of the Stokes profiles (i.e.,
they are the parameters of the Voigt and Faraday-Voigt functions). In
turn, the source-fuction terms, $S_{0}$ and $S_{1}$, control the
continuum level, the line depression, and the Stokes amplitudes.
However, many investigations are based upon the simultaneous ME
inversion of spectral line pairs, like the well-known \ion{Fe}{i}
doublet at 630~nm (e.g., Lites et al. 1993) or the \ion{Mg}{i}~{\em b}
lines at 517.2 and 518.3~nm \citep{1988ApJ...330..493L}. These
inversions are reported to use no extra free parameters as compared to
single-line inversions, based on the similarities between the two
lines belonging to the same multiplet: the opacities for each line are
specified in the ratio of their respective oscillator strengths, while
the Doppler widths and the damping parameters are assumed to be
identical for the two lines. $S_{0}$ and $S_{1}$ are also the same for
both lines since they lie very close in wavelength.

This strategy has been widely used for more than 20 years to analyze
the \ion{Fe}{i} 630 nm measurements taken with instruments such as the
Advanced Stokes Polarimeter \citep{1992SPIE.1746...22E} or the
spectropolarimeter aboard the {\em Hinode} satellite
\citep{2007ASPC..369...55L, 2007SoPh..243....3K, 2008SoPh..249..167T}.
The simultaneous inversion has been shown to provide better results
than single-line inversions \citep[e.g.,][]{1994SoPh..155....1L}. The
better performance is easy to understand: if the observables
reproduced by the inversion are doubled, the results can be expected
to be more accurate (at least by a factor $\sqrt{2}$). However, the
two lines do not have exactly the same atomic parameters (see Table\
\ref{cap5:tablelines}). This makes them to be formed at slightly
different heights, where different physical conditions may exit
\citep[e.g.][]{2006A&A...456.1159M}.  In view of these differences and
of the large excursions of $\eta_{0}$, $a$, and $\Delta
\lambda_\mathrm{D}$ in the real solar photosphere, both horizontally
and vertically, one may wonder whether the simultaneous ME inversion
of the two lines is valid and what the limitations are. General
inversion codes not relying on the ME approximation perform exact line
transfer calculations, so they are able to invert the two lines
without inconsistencies. The purpose of the present Research Note is
to check the ME case using the excellent test bench offered by modern
magnetohydrodynamic (MHD) simulations.

\begin{table}
\caption{\label{cap5:tablelines}Atomic data for the 
\ion{Fe}{i} 630~nm lines. }
\centering
\begin{tabular}{@{}c c c c c c c@{}} \hline $\lambda_{0}$
   (nm) & $\chi$ (eV) & $\log gf$ &  {\scshape Transition }& $\alpha/a^2_0$ & $\sigma$ & $g_\mathrm{eff}$\\ 
\hline \hline 
630.1501 & 3.654 &  $-0.75$  & $5P_2-5D_2$ & 0.243 & 840.5 & 1.67 \\ 
630.2494 & 3.686 &  $-1.236$ & $5P_1-5D_0$ & 0.240 & 856.8 & 2.5 \\ 
\hline \end{tabular}
\tablefoot{
Shown are the central wavelength of 
the transition, $\lambda_{0}$, the excitation potential of the lower
atomic level, $\chi$, the multiplicity of the lower level times the
oscillator strength, $\log gf$, the collisional broadening parameters,
$\alpha$ and $\sigma$ (in units of Bohr's radius $a_0$), and the
effective Land\'e factor of the line, $g_\mathrm{eff}$. The $\log gf$
values have been derived from a fit to the solar spectrum using a 
two-component model of the quiet Sun \citep{borrero2002}.
}
\end{table}


\section{Influence of the ME thermodynamic parameters}
\label{sec:thermodynamic}

Trade-offs among the ME thermodynamic parameters have been reported
and found not to be very important for the inference of vector
magnetic fields and line-of-sight (LOS) velocities
\citep[e.g.,][]{1990ApJ...348..747L,1998ApJ...494..453W}. An explanation 
of this phenomenon has been given by \cite{orozco2007}. Therefore, the 
first question we should answer is whether or not the thermodynamic
parameters of the lines need to be the same to obtain similar LOS
velocities, $v_{\rm LOS}$, magnetic field strengths, $B$,
inclinations, $\gamma$, and azimuths, $\varphi$, from their 
individual analysis.

To investigate this issue, we synthesize realistic Stokes profiles 
for the two \ion{Fe}{i} lines at 630~nm using simulations performed
with the MPS/University of Chicago Radiative MHD code. This code
solves the MHD equations for compressible and partially ionized
plasmas. Further information about the simulations can be found in
\cite{voegler} and \cite{2005A&A...429..335V}. The snapshot
used here belongs to a mixed-polarity simulation run with an average
magnetic field strength of $\langle B \rangle \simeq 140$~G at $\log
\,\tau_{500} =-1$. We generate the Stokes spectra of the two
lines from the MHD models using the SIR code
\citep{1992ApJ...398..375R}, as explained by Orozco Su\'arez et
al.~(2010). We then invert the two lines \emph{separately} with the
MILOS code. The inversion is carried out assuming a one-component
model atmosphere (magnetic filling factor unity) and zero
macroturbulent velocity. No noise is added to the Stokes profiles.

\begin{figure} \centering   
\resizebox{\hsize}{!}{\includegraphics{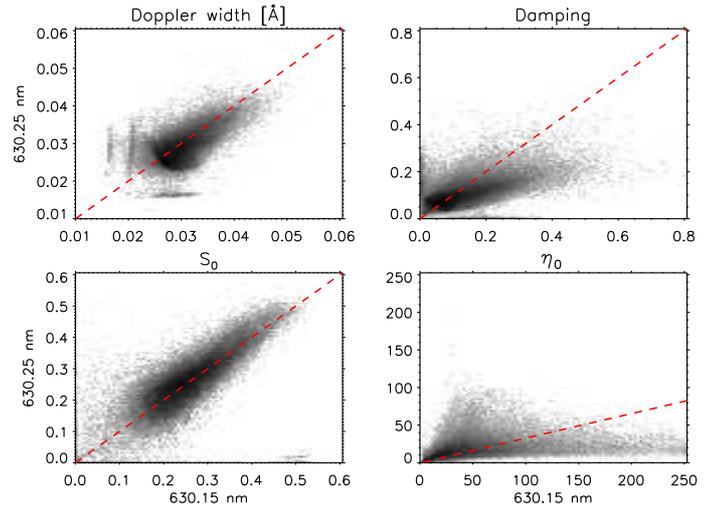}}
\caption{$\Delta\lambda_\mathrm{D}$, $a$, S$_0$, and $\eta_0$ parameters from the inversion of the \ion{Fe}{I} line at 630.25~nm vs those obtained from the 630.15~nm line. The dashed lines represent one-to-one correspondences. In the bottom right panel the dashed line represents $\eta_{0,2}/\eta_{0,1} = 0.327$. Both lines are inverted separately.} \label{Fig1} \end{figure}

Figure~\ref{Fig1} shows scatter plots of
$\Delta\lambda_\mathrm{D}$, $a$, $S_{0}$, and $\eta_{0}$ as inferred
from the inversion of the two lines.  The results for $S_1$ are very
similar to those for $S_0$, with less scatter. In this and other
figures, the gray colors inform about the pixel density, with black
meaning larger values. Over-plotted are dashed lines representing
one-to-one correspondences, except for $\eta_0$ where the ratio
$\eta_{0,2}\,/\,\eta_{0,1}=0.327$ is indicated ($1$ and $2$ stand for
the 630.15 and the 630.25 lines, respectively; see below). The plots
include all the pixels in the simulation snapshot ($288 \times 288$),
independently of the polarization signal. Obviously, the thermodynamic
parameters are far from being the same for the two lines. The scatter
is large for $\Delta\lambda_\mathrm{D}$ and $S_0$, although they show
a linear correlation with a slope close to unity. In the case of
$\eta_0$ and $a$ the scatter is dramatic; differences of up to a
factor 3 for $a$ and of more than one order of magnitude for $\eta_0$
can be seen. As a matter of fact, the $\eta_0$ values obtained for 
the 630.15 nm line span the full range of variation of the 
line-to-continuum opacity ratio in real solar atmospheres 
\citep[see, e.g., Fig.\ 11 of][]{1998ApJ...494..453W}.

\begin{figure} \centering  
\resizebox{\hsize}{!}{\includegraphics{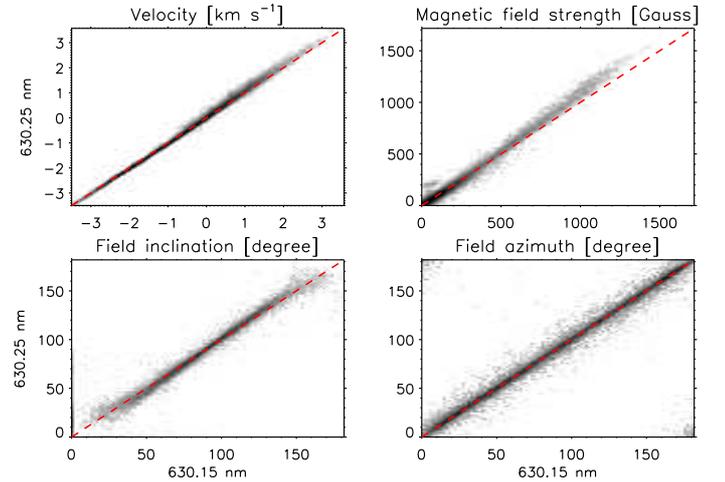}}
\caption{LOS velocity, magnetic field strength, inclination, and azimuth from the inversion of the 630.25~nm line vs those from the inversion of the line at 630.15 nm. Both lines are inverted separately.} \label{Fig2} \end{figure}

Do these discrepancies between the thermodynamic parameters affect the
magnetic and velocity inferences? The answer to this question is
negative as can be seen in Fig.\ \ref{Fig2} where
$v_\mathrm{LOS}$, $B$, $\gamma$, and $\varphi$ are displayed as
inferred from the individual inversion of the 630.15 and 630.25~nm
lines. Despite the strong scatter in the thermodynamics, the results
from both lines are remarkably similar. Therefore, we have to conclude
that although the thermodynamic parameters of ME inversions may have
little meaning, it is possible to establish approximate relations
between the parameters of the two lines for use in simultaneous
inversions, as we shall see in the next section.

\section{Understanding the simultaneous inversion}
\label{MHD}

Since iron is mostly ionized and local thermodynamic equilibrium
conditions prevail in the solar photosphere, the dependence of the
logarithmic derivative of $\eta_{0}$ on temperature (the most
important quantity in line formation) is linear with the excitation
potential $\chi$ and the ionization potential $I$, and quadratic with
the temperature. $\Delta\lambda_\mathrm{D}$ is proportional to the
square root of temperature and does not depend on the atomic
parameters of the line. Assuming van der Waals broadening for the
calculation of the damping coefficient, an intricate (but weak)
dependence of $a$ on $\chi$, along with a proportionality with the
7/10-th power of temperature, is obtained \citep[see,
e.g.,][]{2008oasp.book.....G}.

Neglecting the differences induced by the slightly different
excitation potentials of both lines, the $a$ and
$\Delta\lambda_\mathrm{D}$ ratios are proportional to
$\lambda_{0,2}/\lambda_{0,1}$. Thus, the damping and Doppler width can
safely be assumed equal for both lines, as is also the case for $S_0$
and $S_1$ because the Planck function should not present significant
differences in such a short wavelength interval. $\eta_{0}$ is
proportional to the square of the central wavelength of the line and
to its $gf$ value. Therefore, $\eta_{0,2}/\eta_{0,1} \approx (g_2 \,
f_2 ) / (g_1 \, f_1$), where $g$ is the multiplicity of the lower
level and $f$ the oscillator strength. Note that this ratio is
independent of the temperature or other thermodynamic quantities, so
it should not change with height in the atmosphere even if the opacity
varies by orders of magnitude. Using the parameters of Table 1, the
opacity ratio for the \ion{Fe}{i} 630~nm lines turns out to be
$\eta_{0,2}\,/\,\eta_{0,1} = 0.327$. This is the value implemented in
MILOS. The HAO code and MELANIE use similar ratios\footnote{ The
  opacity ratio depends basically upon $f$, which is known with
  limited precision and varies from one source to the next. This
  causes an uncertainty in the theoretical opacity ratio. For example,
  the laboratory measurements of \citealt{1991A&A...248..315B} give
  $\log gf= -0.718$ for \ion{Fe}{I} 630.15~nm, rather than the $-0.75$
  specified in Table 1. With this oscillator strength, the theoretical
  ratio would be 0.301 (closer to the values retrieved from the
  inversion, but only at the high $\eta_0$ end).}.

\begin{figure}[t] 
\centering 
 \resizebox{\hsize}{!}{\includegraphics{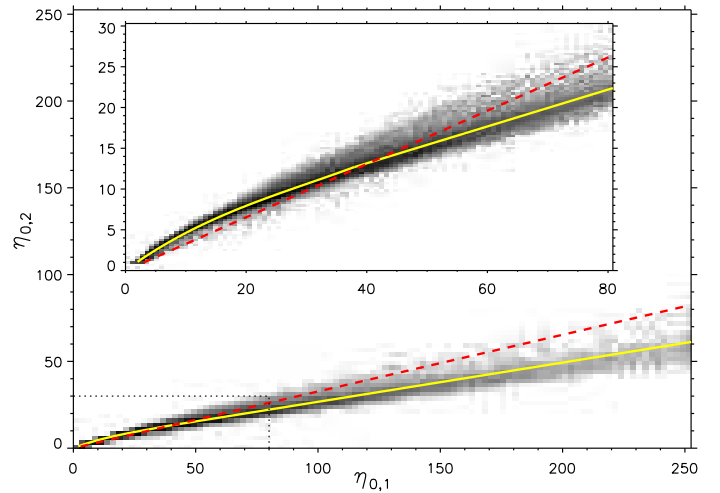}}
 \caption{$\eta_0$ values retrieved from the inversion of the
 \ion{Fe}{I} lines at 630.15 and 630.25~nm. The dashed line represents
 the theoretical ratio $\eta_{0,2}/\eta_{0,1} = 0.327$. The solid
 line stands for a multi-polynomial fit. The inset zooms in the
 dotted box.}
\label{Fig3} 
\end{figure}

To check the validity of this estimate, we invert the 630.25~nm line
again but now forcing $S_0$, $S_1$, $\Delta\lambda_\mathrm{D}$, and
$a$ to be equal to those obtained from the previous inversion of the
630.15~nm line. The remaining model parameters are allowed to vary
freely. Figure \ref{Fig3} shows the $\eta_0$ values retrieved
from the inversion.  The dashed line represents the ratio
$\eta_{0,2}\,/\,\eta_{0,1} = 0.327$. The solid line corresponds to a
multi-polynomial fit $y = a[1]x^3 + a[2]x^2+ a[3]x$ with $a = [ 1
\times10^{-4}, -9.3\times10^{-3}, 0.5]$ for $\eta_{0,1}
\leqslant 38$ and $y = b[1]x + b[2]$ with $b = [ 0.23,4.1]$ for
$\eta_{0,1} > 38$. Note that the theoretical ratio provides a fair
description of the relationship between the two $\eta_{0}$ values in
the range where most of the points are located (to stress the
differences, the inset zooms in on the boxed area). Since the exact
values of the thermodynamical parameters are not very important for
the determination of the magnetic field vector and the LOS velocity,
we conclude that it is safe to use a constant opacity ratio to invert
the two lines simultaneously without increasing the number of free
parameters.

\begin{figure}[!t] \centering 
\resizebox{\hsize}{!}{\includegraphics{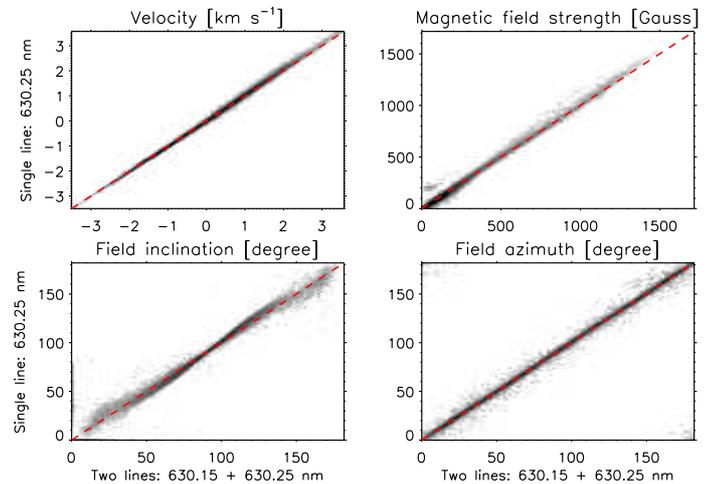}}
\caption{Magnetic field strength, inclination, azimuth, and LOS 
velocity from the the inversion of the 630.25~nm line vs those 
from the simultaneous inversion of the two lines. The latter inversion is done assuming the same model atmosphere (total of 9 free parameters) for the two lines, with their $\eta_{0}$ values coupled.} 
\label{Fig4} 
\end{figure}

The final consistency proof is shown in Fig.\ \ref{Fig4},
where the parameters obtained from the inversion of \ion{Fe}{i}
630.25~nm are plotted against those coming from the simultaneous
inversion of the two lines as coupled through their theoretical
$\eta_0$ ratio. The scatter is very low for $B$ and v$_\mathrm{LOS}$
and somewhat larger for $\gamma$ and $\varphi$, but still much smaller
that that of Fig.\ \ref{Fig2}. 
This suggests that the accuracy of analyses based on one single 
line \citep[e.g.,][]{bommier2009} could be improved by adding the other line.

\begin{acknowledgements}
We thank an anonymous referee for raising the issue investigated in
this paper. Our work has been supported by the Spanish MICINN through
projects AYA2009-14105-C06-06 and PCI2006-A7-0624, by Junta de
Andaluc\'{\i}a through project P07-TEP-2687 (in\-clu\-ding European
FEDER funds), and by the Japan Society for the Promotion of Science.
\end{acknowledgements}


\begin{thebibliography}{}

\bibitem[Auer et al.(1977)]{1977SoPh...55...47A} Auer, L.~H., House, L.~L., 
\& Heasley, J.~N.\ 1977, \solphys, 55, 47

\bibitem[Bard et al.(1991)]{1991A&A...248..315B} Bard, A., Kock, A.,
  \& Kock, M.\ 1991, \aap, 248, 315

\bibitem[Bommier et al.(2009)]{bommier2009} Bommier, V., Mart{\'{\i}}nez Gonz{\'a}lez, M., Bianda, M., Frisch, H., Asensio Ramos, A., Gelly, B., \& Landi Degl'Innocenti, E.\ 2009, \aap, 506, 1415

\bibitem[Borrero \& Bellot Rubio(2002)]{borrero2002} Borrero, J.~M., 
\& Bellot Rubio, L.~R.\ 2002, \aap, 385, 1056 

\bibitem[Borrero et al.(2010)]{2010SoPh..tmp...35B} Borrero, J.~M., 
Tomczyk, S., Kubo, M., Socas-Navarro, H., Schou, J., Couvidat, S., 
\& Bogart, R.\ 2010, \solphys, 35 

\bibitem[Elmore et al.(1992)]{1992SPIE.1746...22E} Elmore, D.~F., et al.\ 
1992, \procspie, 1746, 22

\bibitem[Gray(2008)]{2008oasp.book.....G} Gray, D.~F.\ 2008, The 
Observation and Analysis of Stellar Photospheres.~ Cambridge: 
Cambridge University Press

\bibitem[Harvey et al.(1972)]{1972lfpm.conf..227H} Harvey, J., Livingston, 
W., \& Slaughter, C.\ 1972, Line Formation in the Presence of Magnetic Fields, 227

\bibitem[Kosugi et al.(2007)]{2007SoPh..243....3K} Kosugi, T., et al.\ 
2007, \solphys, 243, 3 

\bibitem[Lagg et 
al.(2004)]{2004A&A...414.1109L} Lagg, A., Woch, J., Krupp, N., \& Solanki, S.~K.\ 2004, \aap, 414, 1109 

\bibitem[Landi Degl'Innocenti 
\& Landolfi(1983)]{1983SoPh...87..221L} Landi Degl'Innocenti, E., \& Landolfi, M.\ 1983, \solphys, 87, 221 

\bibitem[Landolfi \& Landi Degl'Innocenti(1982)]{1982SoPh...78..355L} Landolfi, 
M., \& Landi Degl'Innocenti, E.\ 1982, \solphys, 78, 355 

\bibitem[Landolfi et al.(1984)]{1984SoPh...93..269L} Landolfi, M., Landi 
Degl'Innocenti, E., \& Arena, P.\ 1984, \solphys, 93, 269 

\bibitem[Lites 
\& Skumanich(1990)]{1990ApJ...348..747L} Lites, B.~W., \& Skumanich, A.\ 1990, \apj, 348, 747

\bibitem[Lites et al.(1988)]{1988ApJ...330..493L} Lites, B.~W., Skumanich, 
A., Rees, D.~E., \& Murphy, G.~A.\ 1988, \apj, 330, 493 

\bibitem[Lites et al.(1993)]{1993ApJ...418..928L} Lites, B.~W., Elmore, 
D.~F., Seagraves, P., \& Skumanich, A.~P.\ 1993, \apj, 418, 928 

\bibitem[Lites et al.(1994)]{1994SoPh..155....1L} Lites, B.~W., Martinez 
Pillet, V., \& Skumanich, A.\ 1994, \solphys, 155, 1 

\bibitem[Lites et al.(2007)]{2007ASPC..369...55L} Lites, B.~W., et al.\ 
2007, ASP Conf. Series, 369, 55 

\bibitem[Mart{\'{\i}}nez Gonz{\'a}lez et al.(2006)]{2006A&A...456.1159M} Mart{\'{\i}}nez Gonz{\'a}lez, M.~J., Collados, M., \& Ruiz Cobo, B.\ 2006, \aap, 456, 1159 

\bibitem[Orozco Su{\'a}rez \& Del Toro Iniesta (2007)]{orozco2007} Orozco Su{\'a}rez, D. \& Del Toro Iniesta, J.~C.\ 2007, \aap, 462, 1137 

\bibitem[Orozco Su\'arez et al. (2010)]{Orozco_applicability} Orozco Su\'arez, D., Bellot Rubio, L.R., \& Del Toro Iniesta, J.C.\ 2010, \aap, accepted

\bibitem[Rachkovsky (1962a)]{1962aIzvKrAO...27.148R} Rachkovsky, D.N.\ 1962a, Izv. Krymsk. Astrofiz. Obs., 27, 148

\bibitem[Rachkovsky (1962b)]{1962aIzvKrAO...28.259R} Rachkovsky, D.N.\ 1962a, Izv. Krymsk. Astrofiz. Obs., 28, 259

\bibitem[Ruiz Cobo \& Del Toro Iniesta(1992)]{1992ApJ...398..375R} Ruiz Cobo, B., \& Del Toro Iniesta, J.~C.\ 1992, \apj, 398, 375

\bibitem[Skumanich \& Lites(1987)]{lites2} Skumanich, A., \& Lites, B.~W.\ 1987, ApJ, 322, 473

\bibitem[Socas-Navarro(2001)]{socas2001} Socas-Navarro, H.\ 2001, 
ASP Conf. Series, 236, 487

\bibitem[Tsuneta et al.(2008)]{2008SoPh..249..167T} Tsuneta, S., et al.\ 
2008, \solphys, 249, 167 

\bibitem[Unno(1956)]{1956PASJ....8..108U} Unno, W.\ 1956, \pasj, 8, 108

\bibitem[V{\"o}gler(2003)]{voegler} V{\"o}gler, A.\ 2003, PhD Thesis, University of 
G{\"o}ttingen, Germany, http://webdoc.sub.gwdg.de/diss/2004/voegler/ 

\bibitem[V{\"o}gler et al.(2005)]{2005A&A...429..335V} V{\"o}gler, A., Shelyag, S., 
Sch{\"u}ssler, M., Cattaneo, F., Emonet, T., \& Linde, T.\ 2005, A\&A, 429, 335 

\bibitem[Westendorp Plaza et al.(1998)]{1998ApJ...494..453W} Westendorp Plaza, C., Del 
Toro Iniesta, J.~C., Ruiz Cobo, B., Mart\'inez Pillet, V., Lites, B.~W., \& Skumanich, 
A.\ 1998, \apj, 494, 453 

\end{thebibliography}
\end{document}